\RequirePackage{fix-cm}
\documentclass[smallextended,UKenglish]{svjour3}       
\smartqed  
\usepackage{graphicx}
\usepackage[utf8]{inputenc}
\usepackage{natbib}
\usepackage{amssymb}
\usepackage{float}
\usepackage{hyperref}

\usepackage{amsmath,amssymb,amsfonts,bbm}
\usepackage{dsfont} 
\usepackage[svgnames]{xcolor}
\usepackage{graphicx}

\usepackage{algorithm2e,algpseudocode}
\usepackage{array}
\usepackage{indentfirst}

\usepackage{url}
\usepackage{soul}

\usepackage{algorithm2e,algpseudocode}
\usepackage{array}
\usepackage{indentfirst}

\usepackage{booktabs} 
\usepackage{etoc} 
\usepackage{multirow}
\usepackage{makecell}
\usepackage{enumitem}

\usepackage{subcaption}

\usepackage{soul}
\usepackage[colorinlistoftodos,prependcaption,textsize=tiny]{todonotes}   

\definecolor{navy}{rgb}{0.1, 0.1, 0.8}
\definecolor{gray}{rgb}{0.6, 0.6, 0.6}
\definecolor{myblue}{rgb}{.8, .8, 1}
\definecolor{olive}{rgb}{0.1, 0.5, 0.1}
\definecolor{ao}{rgb}{0.0, 0.5, 0.0}

\newcommand{\eat}[1]{}

\usepackage{xspace}

\newcommand{\ed}{\text{ed}}
\newcommand{\sed}{\text{sed}}
\newcommand{\dist}{\text{dist}}
\newcommand{\simt}{\text{sim}}

\newcommand{\del}{\text{del}}
\newcommand{\ins}{\text{ins}}
\newcommand{\sub}{\text{sub}}

\usepackage{amstext}
\usepackage{tikz}
\usetikzlibrary{arrows,decorations.pathmorphing,fit,positioning}
\usepackage{babel}
\usepackage[font=small,labelfont=bf]{caption}

\begin{document}

\title{Analysing user identity via time-sensitive semantic edit distance (t-SED): A case study of Russian trolls on Twitter
}

\titlerunning{User identity via time-sensitive semantic edit distance}        

\author{Dongwoo Kim \and Timothy Graham \and Zimin Wan \and Marian-Andrei Rizoiu}


\institute{Dongwoo Kim \at
              Australian National University \\
              \email{dongwoo.kim@anu.edu.au}           
           \and
           Timothy Graham \at
              Queensland University of Technology \\
              \email{timothy.graham@qut.edu.au} 
          \and
          Zimin Wan \at
              Australian National University \\
              \email{u6013849@anu.edu.au} 
          \and
          Marian-Andrei Rizoiu \at
            University of Technology Sydney \\
            \email{Marian-Andrei.Rizoiu@uts.edu.au}
}

\date{Received: date / Accepted: date}

\maketitle

\begin{abstract}
In the digital era, individuals are increasingly profiled and grouped based on the traces they leave behind in online social networks such as Twitter and Facebook. 
In this paper we develop and evaluate a novel text analysis approach for studying user identity and social roles by redefining identity as a sequence of timestamped items (e.g. tweet texts). 
We operationalise this idea by developing a novel text distance metric, the \textit{time-sensitive semantic edit distance} (t-SED), which accounts for the temporal context across multiple traces. 
To evaluate this method we undertake a case study of Russian online-troll activity within US political discourse. 
The novel metric allows us to classify the social roles of trolls based on their traces, in this case tweets, into one of the predefined categories left-leaning, right-leaning, and news feed.
We show the effectiveness of the t-SED metric to measure the similarities between tweets while accounting for the temporal context, and we use novel data visualisation techniques and qualitative analysis to uncover new empirical insights into Russian troll activity that have not been identified in previous work. 
Additionally, we highlight a connection with the field of Actor-Network Theory and the related hypotheses of Gabriel Tarde, and we discuss how social sequence analysis using t-SED may provide new avenues for tackling a longstanding problem in social theory: how to analyse society without separating reality into micro versus macro levels.
\end{abstract}

\section{Introduction}

Over the past decade, social media platforms such as Twitter and Reddit have exploded in popularity, offering a public forum for millions of people to interact with each other. 
The availability of social media data through Application Programming Interfaces (APIs) has provided unprecedented opportunities for social research,
particularly concerning learning about and defining user identities through the digital traces that actors leave behind~\citep{latour2012whole}.
In this work, we consider that individuals' online identities are defined by their digital traces, and that individuals who are similar to each other should have similar sequences of trace activity.
On Twitter, such a trace would be the sequence of their authored tweets, and two sequences are considered similar when transforming one sequence into another (by adding, deleting or substituting elements) requires few operations. 
In other words, the `edit distance' between two sequences should be small if two traces are similar. 
This ideas extends from what \citet{abbott1995sequence} has termed as sequence analysis in social science, which has been more recently defined as `social sequence analysis' \citep{cornwell2015social}. 
Originally, optimal matching (OM) or string matching methods were developed to efficiently analyse protein and DNA sequences at scale.
The aim was to discover close matches to a particular sequence of interest such as a newly identified protein \citep{abbott2000sequence}. 
Social sequence analysis has been used to study topics such as career trajectories, daily life, and national histories. 
The goal is usually to discover interesting patterns within sequential datasets, by using an edit distance algorithm to compute the pairwise distances between pairs of sequences, and analysing the resulting distance matrix using a clustering or dimensionality reduction technique. 
To our knowledge the present study is the first to propose the use of sequence analysis to study identity via trace text data. 

To measure the similarity between two texts (e.g., tweets), we propose the \emph{time-sensitive semantic edit distance} (t-SED), a novel variant of edit distance adapted to natural language by embedding two factors: \emph{word similarity} and \emph{time sensitivity}. As we argue later on, semantics and temporal information are important, and a naive application of conventional edit distance would lose this information. 
The word similarity modulates the cost of the edit operations (i.e. deletion, insertion and substitution) according to the similarity of the words involved in the operation.
The time sensitivity is highest when the two tweets are concomitant (i.e. they were authored within a short time interval of each other), and addresses issues such as changing discussion topics and concept drift.

To provide an empirical grounding for this work, we undertake a case study analysis of a well-known problem of current interest: Russian troll activity before and after the 2016 US Election.
Online trolls are predominantly human or hybrid (semi-automated) user accounts who behave in a deceptive, destructive, and/or disruptive manner in a social setting on the Internet \citep{BUCKELS201497}.
Social bots are largely automated systems that pose as humans, and which seek to influence human communication and manipulate public opinion at scale. 
Bots have recently caused controversy during the 2016 U.S. presidential election, when it was found that they were not only highly prevalent but also highly influential and ideologically driven \citep{FM7090,rizoiu2018debatenight,Kollanyi.2016.presidentialdebate}. 
The troll and bot account sets are not necessarily mutually exclusive, and recent studies uncovered that the Russian interference during the 2016 U.S. Election involved a combination of both types of accounts \citep{ferrara-et-al-trolls-ideology} to weaponise social media, to spread state-sponsored propaganda and to destabilise foreign politics \citep{broniatowski-et-al,ICWSM1613006}. 

There are several challenges that arise when studying the actions of malicious actors such as trolls and bots.
The first challenge concerns distinguishing between their different types (or roles) in order to understand their strategy.
Current state-of-the-art approaches usually build large amounts of features, and they use supervised machine learning methods to detect whether a Twitter account exhibits similarity to the known characteristics of social bots~\citep{davis2016botornot}, and use text mining and supervised classification methods to identify online trolls~\citep{mihaylov2015finding}.
However, such approaches have several drawbacks, including requiring access to (often private) user features, and periodic retraining of models to maintain up-to-date information~\citep{mihaylov2015finding}.
The question is can we circumvent this arms race of distinguishing the roles of online trolls? Can we develop sociologically-grounded approaches to analysing identity and social position that do not over-rely on machine learning algorithms, but instead utilise sequence analysis?
The second challenge lies in analysing and understanding the strategy of trolls.
While prior work has focused on troll detection, recent studies show the existence of sub-types of trolls simulating multiple political ideologies~\citep{boatwrighttroll,ferrara-et-al-trolls-ideology,2018who-let-the-trolls-out,stewart2018examining}.
This suggests a sophisticated and coordinated interplay between the different types of troll behaviour to manipulate public opinion effectively.
The empirical question is therefore how can we use sequence analysis to understand the behaviour and the strategy of trolls over time, and analyse the interplay between different troll sub-roles?

This paper addresses the above challenges using a publicly available dataset of Russian troll activity on Twitter, published by \citet{boatwrighttroll}, consisting of nearly 3 million tweets from 2,848 Twitter handles associated with the Internet Research Agency -- a Russian ``troll factory'' -- between February 2012 and May 2018.

We tackle the problem of distinguishing the roles and identities of Russian trolls on Twitter by operationalizing the methodology of sequence analysis \citep{abbott1995sequence}: we measure the similarities between the sequence of texts they generate over time against a reference set. In doing so this provides a form of `direct validation' for the results \cite[p.~150]{abbott2000sequence}. 
We show that for the task of distinguishing troll roles based on ground truth labels (e.g., left versus right trolls), our method outperforms a logistic regression baseline learner by more than $36\%$ (macro F1).

We address the second challenge by constructing a two-dimensional visualisation, in which we embed the traces of Twitter trolls and their similarities (measured using t-SED) by using t-SNE~\cite{maaten2008visualizing}.
In addition to observing that the tweets emitted by trolls with similar roles cluster together -- which was expected from the operationalisation of sequence analysis -- we make several new and important findings:
(1) despite trolls having different and distinguishable roles, they worked together: the right trolls impersonate a homogeneous conservative identity, while the left trolls surround the conversation from all sides, with messages that simultaneously divide the Democrat voters on key issues and complement the destabilisation strategies of the right trolls;
(2) we observe clusters of complex and orchestrated interplay between left and right trolls, both attempting to co-opt and strategically utilise ethnic identities and racial politics (e.g. \textit{Black Lives Matter}), as well as religious beliefs;
(3) the news feed trolls (i.e. impersonating news aggregators) are observed to have multiple agendas and sub-roles, such as clusters that disproportionately tweet about news of violence and civil unrest to create an atmosphere of fear, and other clusters that exhibit politically-biased reporting of federal politics; and
(4) there is an obvious shift of strategy between before and after the elections, with less distinguishable roles and strategies of trolls after the elections. 

Beyond the application to studying Russian trolls on Twitter, we believe this work has important implications for computational social science and social theory. 
We show that a sequence analysis approach can not only accurately recover the ground-truth labels of dynamic social data (in this case tweets authored by different types of Russian trolls), but it can also reveal new insights and sensical results. 
Hence this approach not only recovers the identity of social actors, but more importantly it `rediscovers' their identity by tracking and quantifying their trace associations (i.e., distances via t-SED) with other actors over time. 
Recent developments in Actor-Network Theory (ANT) suggest that these digital traces may provide an empirical basis to fundamentally reevaluate how we define social action and actors, building on ideas originally set out by Gabriel Tarde in the 19$^{th}$ century \citep{latour2002gabriel,latour2012whole}. To date, little progress has been made to offer a computational solution. We believe this paper takes some important first steps towards advancing and formalising this body of theory within computational social science.

\paragraph{To summarise, the main contributions of this work include:}
\begin{itemize}
	\item We introduce a \textbf{sociologically-grounded classification framework to identify the identity and social role of online trolls} using trace data;
	\item We operationalise \textbf{social sequence analysis} as a way to measure identity;
	\item We develop a \textbf{time-sensitive semantic edit distance} (t-SED) to measure similarity between two elements that form part of traces (i.e. tweets), embedding two components of online natural language: word similarity and item concomitance;
	\item We propose \textbf{a visualisation based on the novel distance metric} to gain new insights into the strategic behaviour of trolls before and after the 2016 U.S. presidential election; and
	\item We take several first steps towards bridging computational social science and Actor-Network Theory.
\end{itemize}

The remainder of this paper is organised as follows. In Section \ref{sec:background-related}, we turn attention to the empirical problem of studying political troll activity on social media, and link this back to the research problem of measuring identity and social position using sequence analysis. In Section \ref{sec:troll-dataset} we provide details of the data used in the empirical analysis. In Section \ref{sec:sed} we operationalise sequence analysis with the aim of distinguishing patterns in troll identities and activities. Based on this we introduce a new time-sensitive distance measure suited for texts and a majority voting framework for the analysis. We then set out t-SED, a modified edit distance algorithm. In Section \ref{sec:results-findings} we measure the classification performance of our proposed method on the Russian troll dataset. In Section \ref{sec:visualisation}, an analysis of the results is provided, with specific emphasis on qualitative interpretation of the clusters that emerge from the sequence analysis. Following this, Section \ref{sec:discussion} undertakes a discussion of the results and their implications for a longstanding problem in social theory. The paper concludes with a summary of the main findings and contributions, along with the limitations of the research and possible directions for future work. 

\section{Background and related work}
\label{sec:background-related}

\subsection{Political trolls on social media}
\label{subsec:political-trolls-detection}

\textbf{Bots and trolls in political discussions.}
In recent years online trolls and social bots have attracted considerable scholarly attention. 
Online trolls tend to be either human, `sock puppets' controlled by humans \citep{Kumar-et-al}, or semi-automated accounts that provoke and draw people into arguments or simply occupy their attention \citep{herring2002searching} for amplify particular messages and manipulate discussions \citep{broniatowski-et-al,ferrara-et-al-trolls-ideology}.
Recent studies have investigated the impact of trolls and bots in social media to influence political discussions \citep{FM7090}, spread fake news \citep{shao2017spread}, and affect the finance and stock market \citep{ferrara2016rise}.
Especially, in a political context, studies have shown that online trolls mobilised support for Donald Trump's 2016 U.S. Presidential campaign \citep{ICWSM1613006}, and, of particular interest to this paper, were weaponised as tools of foreign interference by Russia during and after the 2016 U.S. election \citep{boatwrighttroll,2018who-let-the-trolls-out}. 
It is the current understanding that Russian trolls successfully amplified a largely pro-Trump, conservative political agenda during the 2016 U.S. Election, and managed to attract both bots and predominantly conservative human Twitter users as `spreaders' of their content \citep{ferrara-et-al-trolls-ideology,stewart2018examining}. 

\textbf{Detection and role of trolls.}
While trolls and bots have become increasingly prevalent and influential, methods to detect and analyse their role in social networks has also received wide attention \citep{cook2014twitter,davis2016botornot,varol2017online}.
On the other hand, identifying and differentiating specific sub-groups or types of trolls poses a difficult challenge, which has attracted relatively less attention.
Different types of trolls can be employed by specific individuals or groups to achieve specialised goals, such as trolls employed by the Internet Research Agency (IRA) to influence the political discourse and public sentiment in the United States \citep{boatwrighttroll,stewart2018examining}. 
The Clemson researchers \cite{boatwrighttroll} used advanced tracking software of social media to collect tweets from a large number of accounts that Twitter has acknowledged as being related with the IRA. Using qualitative methods, the researchers identified five types of trolls, namely right troll, left troll, news feed, hashtag gamer, and fearmonger \citep{boatwrighttroll}. They found that each type of troll exhibited vastly different behaviour in terms of tweet content, reacted differently to external events, and had different patterns of activity frequency and volume over time \citep[p.10-11]{boatwrighttroll}. 

\textbf{Relation to our work.}
Therefore we observe a close connection between \textit{semantics} (what trolls talk about), \textit{temporality} (when they are active and how hashtags are deployed over time), and the particular \textit{roles} and strategies that drive their behaviour (e.g. right versus left troll). 
From an analytical point of view, our interest in this paper is to develop and evaluate a framework that can accurately identify roles of users within a sub-population (in this case Russian troll types) by clustering them based not only on semantics but also temporality, that is, the order in which activities occur over time. To our knowledge this has not been achieved in previous work, where temporal information is often ignored or disregarded in analysis, such as the use of cosine distance and Levenshtein edit distance to differentiate sockpuppet types \citep{Kumar-et-al}, network-based methods to infer the political ideology of troll accounts \citep{ferrara-et-al-trolls-ideology}, or word embeddings and hashtag networks with cosine distance as edge weights \citep{2018who-let-the-trolls-out}. Where temporal analysis of troll activity has been undertaken, the focus has been on tweet volume and hashtag frequency at different time points \citep{2018who-let-the-trolls-out} rather than how roles and strategies change over time. 


\section{Russian trolls dataset}
\label{sec:troll-dataset}

This study uses a publicly available dataset of verified Russian troll activity on Twitter, published by Clemson University researchers \citep{boatwrighttroll}\footnote{Available at https://github.com/fivethirtyeight/russian-troll-tweets/}. The Russian troll tweets dataset consists of nearly 3 million tweets from 2,848 Twitter handles associated with the Internet Research Agency, a Russian ``troll factory''. It is considered to be the most comprehensive empirical record of Russian troll activity on social media to date. 
The tweets in this dataset were posted between February 2012 and May 2018, most between 2015 and 2017. Note that the dataset is constructed by using handles provided by Twitter to the House Intelligence Committee, which may not reflect the entire population of Russian trolls.

In this work, we aim to distinguish trolls based solely on their authored text.
We focus on the tweet content, and we try to detect the author category for each tweet (i.e. what type of Russian troll it was authored by), rather than detect the author category for each handle. 
There are multiple types of Russian trolls categorised by the Clemson researchers, namely: right troll; news feed; left troll; hashtag gamer; and fearmonger, sorted in decreasing order by tweet frequency. 
In this research, we will focus on the top 3 most frequent trolls: right troll, news feed and left troll. In addition, in this study we only consider English-language tweets, although future work can easily generalise our methods to any language expressed as Unicode. 
This amounts to a subset of 1,737,210 tweets emitted by 733 accounts.
We have removed the hashtag gamer category in the further analysis since the identity of these accounts are easily identified due to the unique usage of hashtags. The fearmonger category has been removed due to the sparsity of the category.

According to \citet{boatwrighttroll}, right trolls behave like ``MAGA\footnote{MAGA is an acronym that stands for Make America Great Again. It was the election slogan used by Donald Trump during his election campaign in 2016, and has subsequently become a central theme of his presidency.} 
Americans'', they mimic typical Trump supporters and they are highly political in their tweet activity. 
On the other hand, left trolls characteristically attempt to divide the Democratic Party against itself and contribute to lower voter turnout. 
They achieve this by posing as mimic Black Lives Matter activists, expressing support for Bernie Sanders\footnote{Bernie Sanders was the alternative Democrat Presidential Nominee}, and acting derisively towards Hillary Clinton. 
While tweets posted by left and right trolls have a strong political inclination, news feeds trolls tend to present themselves as legitimate local news aggregators with the goal of contributing to, and magnifying, public panic and disorder.

\section{Classifying the roles of trolls}
\label{sec:sed}

\subsection{Time-sensitive metric for trace classification}
\label{subsec:majority-voting-KNN}

Given that each tweet is labelled using one target label, our goal aims to distinguish types of accounts based on their tweets.
Unlike a general supervised classification problem, where pairs of an input and label are provided as an example, our classification algorithm needs to predict a label from a set of timestamped text snippets authored by a target account. 
We formulate the user account classification as a majority voting problem, where the majority label of tweets is the predicted label of the corresponding account. 
The open question is how to classify the label of an individual tweet given a set of training traces (i.e. sequences of tweets) in terms of the time-sensitive similarity.

We employ the $k$-nearest neighbour algorithm (KNN) to predict the labels of tweets. 
A KNN classifies a data point based on the majority labels of $k$-nearest neighbour based on \textit{distances} to the other labelled data points.
Let $\boldsymbol{s}_i$ and $\boldsymbol{s}_j$ be two sequences of tokens, constructed by tokenising tweets $i$ and $j$;
let $t_{i}$ and $t_{j}$ be the timestamps of tweets $i$ and $j$, respectively. 
We propose a time-sensitive distance metric between tweets $i$ and $j$ formulated as
\begin{align}
    D(i, j) = \dist(\boldsymbol{s}_i, \boldsymbol{s}_j) \times \exp(\theta |t_{i} - t_{j}|),\quad \theta>0
    \label{eqn:tsknn}
\end{align}
where $\dist(\boldsymbol{s}_i, \boldsymbol{s}_j)$ measures a distance between tokenised sequences $\boldsymbol{s}_i$ and $\boldsymbol{s}_j$ and the exponential term 
penalises large timestamp differences between the two tweets.
This is required because sometimes seemingly similar text snippets may represent completely different concepts, because the meaning of the employed terms has evolved with time.
For instance, the hashtag \#MeToo had a general meaning prior to 15 October 2017, whereas afterwards the meaning of the same hashtag changed dramatically with the emergence of the \#MeToo social movement on Twitter.
By adopting the exponential penalisation inspired from point process theory~\cite{leskovec2008microscopic,rizoiu2018hawkes}, the KNN weights more towards temporally related tweets.

In general, the Euclidean distance metric is employed for the function $\dist(\boldsymbol{s}_i, \boldsymbol{s}_j)$, when $\boldsymbol{s}_i$ and $\boldsymbol{s}_j$ are defined in an Euclidean space. 
However, in our case, $\boldsymbol{s}_i$ and $\boldsymbol{s}_j$ are sequences of word tokens for which the Euclidean distance is not defined. 
One may use a bag-of-words representation of tokens to map the sequence of tokens into a vector space, and then employ a text distance metric such as cosine distance.
However, the bag-of-words representation loses the ordering of tokens, which may embed thematic concepts that cannot be understood from individual words.
In the next section, we propose a new distance metric to compute the distance between two sequences of tokens while preserving the temporal ordering between tokens. 

\subsection{Semantic edit distance}
\label{subsec:semantic-edit-distance}

We propose a new text distance metric $\dist(\cdot, \cdot)$ to capture semantic distance between two sequence of symbols based on the edit distance (ED) metric. 
The ED is a method to quantify the distance between two symbolic sequences by calculating the minimum value of the required edit operations to transform one sequence into the other. 
There may be different sets of operations according to the definition of ED. 
The most common form of ED is known as Levenshtein distance~\cite{navarro2001guided}. 
In Levenshtein's original definition, the edit operations include the insertion, deletion and substitution, and each of these operations has a unit cost. 
Therefore the original ED is equal to the minimum number of the required operations to transform one string into the other. 

Formally, given two sequences $\boldsymbol{a} = a_1, a_2, ..., a_n$ and $\boldsymbol{b} = b_1, b_2, ..., b_m$, the edit distance is the minimum cost of editing operations required to transform $\boldsymbol{a}$ into $\boldsymbol{b}$ via three operations: (i) insert a single symbol into a string; (ii) delete a single symbol from a string and (iii) replace a single symbol of a string by another single symbol, associated with non-negative weight cost $w_{\ins} (x)$, $w_{\del} (x)$ and $w_{\sub} (x,y)$, respectively. 
Let $\boldsymbol{a}$ and $\boldsymbol{b}$ be sequences of $n$ and $m$ symbols, respectively. The edit distance between $\boldsymbol{a}$ and $\boldsymbol{b}$ is given by $\ed(n,m)$, defined recursively,
\begin{align*}
  \ed(i,0) =& \sum\limits_{k=1}^i w_{\del} (a_k) \hspace{6.5em} \text{for } 1 \leq i \leq n\\
  \ed(0,j) =& \sum\limits_{k=1}^j w_{\ins} (b_k) \hspace{6.4em} \text{for } 1 \leq j \leq m
\end{align*}
\begin{align*}  
  \ed(i,j) =&
  \begin{cases}
    \ed(i-1,j-1) \hfill \text{if } a_i = b_j\\
    \min 
    \begin{cases}
        \ed(i-1,j) + w_{\del} (a_i)\\
        \ed(i,j-1) + w_{\ins} (b_j)\\
        \ed(i-1,j-1) + w_{\sub} (a_i,b_j)
    \end{cases}
    \\ \hfill \text{otherwise.}
  \end{cases}
\end{align*}
In the original ED, each operation is often assumed to have a unit cost. The unit cost assumption is convenient to measure the distances, however, it does not reflect the similarity between different symbols.
For example, given sentences $\boldsymbol{s}_1 =$ ``I like music'', $\boldsymbol{s}_2 =$ ``I love music'' and $\boldsymbol{s}_3 =$ ``I hate music'', $\ed(\boldsymbol{s}_1, \boldsymbol{s}_2) = \ed(\boldsymbol{s}_2, \boldsymbol{s}_3) = 1$ if we assume word level edit distance where each symbol corresponds to a word token in a vocabulary. However, ``love'' and ``like'' have more similar meaning than ``love'' and ``hate'' or ``like'' and ``hate''. 
In this situation, we expect that the distance between $\boldsymbol{s}_1$ and $\boldsymbol{s}_2$ should be less than the distance between $\boldsymbol{s}_2$ and $\boldsymbol{s}_3$.

As it turns out from the previous example, it is essential to understand and measure similarity between word symbols to further measure the distance between sentences.
No canonical way exists to measure similarities between words, but it is often assumed that if words are used frequently in similar context, then the words play a similar role in sentences.
To capture the context of words, we compute the co-occurrence statistics between words, which shows how often a certain word is used together with the other words.
We first construct a co-occurrence matrix based on the number of times a pair of words is used in the same tweet. 
Then, to measure how frequently a pair of words are used in similar context, we compute the cosine similarity between two co-occurrence vector corresponding to the rows from the co-occurrence matrix. 
Therefore, the more two words are used frequently in a similar context provided by co-occurring words, the more similar they are based on the cosine similarity. 
From now on, we denote $\simt(x,y)$ as the cosine similarity of word $x$ and $y$ from the co-occurrence matrix.
Many previous studies have shown that the co-occurrence patterns can capture meaningful properties of words including synonymousness~\cite{mikolov2013efficient,pennington2014glove}.

We propose an edit distance endowed with novel cost functions of the three edit operations, named semantic edit distance (SED), using the word similarity.
The more similar two sentences are, the fewer the editing operation should cost. In other words, the cost of operation equals to the dissimilarity of two words. Based on this intuition, we propose three cost functions for edit operations as follows:
\begin{align} 
\label{eqn:del}
  w_{\del} (a_i) &= 1 - \simt(a_i, a_{i-1})\\
\label{eqn:ins}
  w_{\ins} (b_i) &= 1 - \simt(b_i, b_{i-1})\\
\label{eqn:sub}
  w_{\sub} (a_i,b_j) &= 1 - \simt(a_i, b_j)
\end{align}
The intuitions behind each of cost function are
\begin{itemize}
  \item For the deletion in \autoref{eqn:del} (or insertion in \autoref{eqn:ins}, resp.), if two consecutive symbols are similar, deleting (inserting) the latter one would not cost much, and the deletion (insertion) operation would have little influence on distance between two strings. 
  \item For the substitution in \autoref{eqn:sub}, if the symbol $a_i$ of sequence $\boldsymbol{a}$ is similar to the symbol $b_j$ of sequence $\boldsymbol{b}$, the substitution should not cost much. 
\end{itemize}
We denote $\sed(\cdot, \cdot)$ as $\ed(\cdot, \cdot)$ endowed with the above three operation costs.
Finally, applying $\sed(\cdot, \cdot)$ to $\dist(\cdot, \cdot)$ in \autoref{eqn:tsknn} results in a time-sensitive semantic edit distance, denoted as t-SED in the rest of this paper.



\section{Evaluation of role prediction}
\label{sec:results-findings}
In this section, we measure the classification performance of our proposed method on the Russian troll dataset.


\paragraph{Experimental settings.}

\begin{table}[t!]
\centering
\begin{tabular}{crrr}
    \toprule
     & Train & Validation & Test \\ \midrule
    Left troll & 101 & 53 & 79 \\ 
    Right troll & 186 & 106 & 155 \\ 
    News feed & 20 & 11 & 22 \\
    \bottomrule
\end{tabular}
\caption{Number of accounts for each category used for training, validation, and testing. 50 tweets are sampled for each account in the classification experiments.}
\label{tab:dataset}
\end{table}

We use a subset of the Russian troll dataset consisting of users labelled as right trolls, left trolls and news feed, as described in \autoref{sec:troll-dataset}. 
We split the accounts into 50\% train, 20\% validation, and 30\% test datasets. 
The detail statistics of each dataset is described in \autoref{tab:dataset}. 
For each account, we randomly sample 50 tweets for ease of computation. 
We tokenise the text of tweets using a tweet pre-processing toolkit\footnote{Available at https://github.com/s/preprocessor} and remove infrequent words which occur less than 3 times in the corpus.
The co-occurrence matrix used to compute the word similarity is constructed from the entire dataset. 

We test the proposed KNN approach with three distance measures (i.e. cosine distance\footnote{The bag-of-words is used to map a sequence to vector}, ED and SED), and their time-sensitive counterparts as defined in \autoref{eqn:tsknn} (denoted as t-Cosine, t-ED and t-SED, respectively). 
Note that the SED ranges from zero to the maximum length of sentences, therefore SED depends on the lengths of sequences; short sentences are likely to have small SED. 
To investigate the effect of sequence length, we additionally propose and test two normalised SEDs: SED/Max\footnote{Length normalisation: $\sed(\boldsymbol{a},\boldsymbol{b})/\max(\boldsymbol{|a|},\boldsymbol{|b|})$} and SED/ED\footnote{Ratio normalisation: $\sed(\boldsymbol{a},\boldsymbol{b})/\ed(\boldsymbol{a},\boldsymbol{b})$}.
As a baseline, we also test a logistic regression classifier
with and without temporal information (denoted as LR and t-LR). 
We train the LR models with bag-of-words features to classify the label of an individual tweet and predict account label based on majority vote. 
To add temporal information into the logistic regression, we compute the normalised timestamp of tweets and we add it into the feature set.
Finally, the classification performance is measured by macro and micro F1. 
The F1 score is the geometric mean of the precision and the recall of the classification; to obtain a high F1 score, both the precision and the recall need to be high.
The macro F1 score is particularly useful for datasets with a skewed distribution between the classes -- note that our dataset shows highly skewed distribution toward the right troll accounts.
Not predicting correctly the minority class penalizes severely the macro F1.

\autoref{tab:test_perf} summarises all the tested approaches and their performances.

\begin{figure}[t!]
    \centering
    \includegraphics[width=\linewidth]{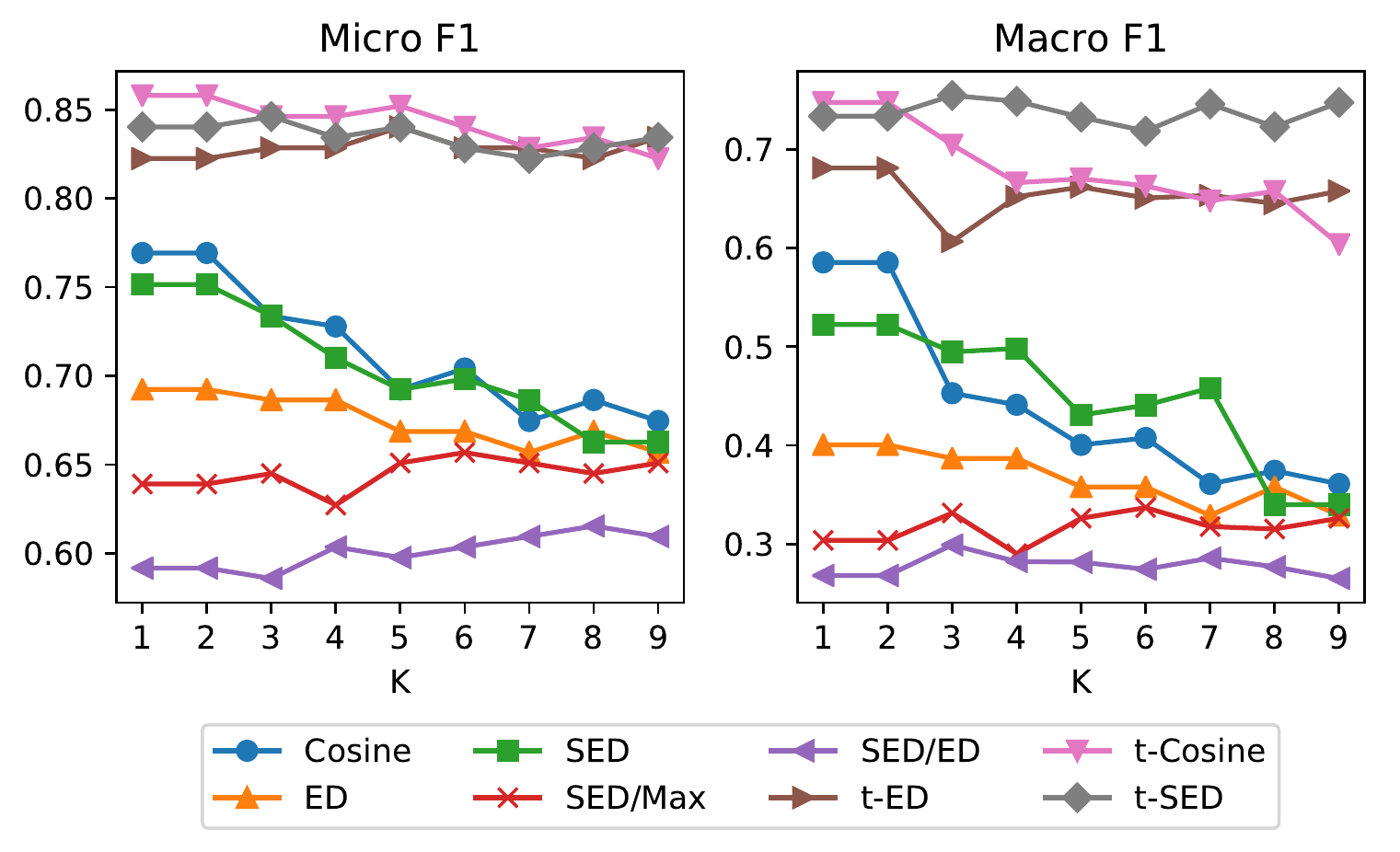}
    \caption{Macro and micro F1 scores on validation set with KNN. Cosine, ED, and SED performs the best when k is 1. t-SED shows relatively consistent performances over varying numbers of neighbours.}
    \label{fig:valid_perf}
\end{figure}

\begin{table}[t!]
    \centering
    \caption{Micro and macro F1 scores on test set along with the number of neighbours (K) for KNN. Although SED outperforms all baseline models, t-SED significantly outperform their non time-sensitive counterparts implying the importance of incorporating the temporal dimension.}
    \begin{tabular}{llrrrr}
        \toprule
        \multicolumn{2}{c}{}& \multicolumn{2}{c}{Micro F1} &  \multicolumn{2}{c}{Macro F1} \\ 
        & Method & K & F1 & K & F1\\ \midrule
    
        \multirow{3}{*}{Baseline}&LR & - & 0.75 &- &0.55\\
        & ED & 1 &0.73 & 1 & 0.47\\
        &Cosine & 1 & 0.75 & 1 & 0.54 \\ \midrule
    
        \multirow{3}{*}{Semantic}&SED	&1	&0.79&1	&0.62\\
        &SED/Max &6	&0.68&1	&0.39\\
        &SED/ED	& 8 &0.62&8	&0.34\\ \midrule
        
        \multirow{4}{*}{Temporal}&t-LR	& - & 0.79& -& 0.61\\
        &t-ED	& 1 & 0.84& 1& {0.76}\\
        &t-Cosine	&5	&0.81&1	&0.61\\
        &t-SED & 3 & \textbf{0.86} & 3 & \textbf{0.78} \\
    \bottomrule
    \end{tabular}

    \label{tab:test_perf}
\end{table}

\paragraph{Results.}
\label{sec:result-section}

\autoref{fig:valid_perf} shows the classification performances of different metrics on the validation dataset with varying number of neighbourhood size in KNN. Note that t-SED has additional parameter $\theta$, which controls the exponential rate. We perform a grid search on $\theta$ to find the best configuration and report the best validation performance in \autoref{fig:valid_perf}. One interesting pattern from the validation set is that all other metrics except time sensitive metrics suffer from having a large number of neighbours, whereas the time sensitive metrics retain stable performances across the different number of neighbours.
We conjecture that including more neighbours include more tweets from different time ranges, as done with the non time-sensitive metrics, eventually hurts the classification of individual tweets.

The results shown in \autoref{tab:test_perf} are the best performances of each approach over the range of $k$ on the validation set.
Overall, the t-SED outperforms all other metrics for both macro and micro F1 scores. 
We interpret the results from four perspectives: 
\textbf{1) The importance of word similarity} by comparing ED and SED.
By adding the word similarity into the cost function, SED can significantly outperform ED in the role prediction.
\textbf{2) The importance of preserving order between tokens in tweets.} 
Although the naive ED performs worse than the cosine distance, SED outperforms cosine
outlines the importance of preserving order between tokens, alongside with accounting for word similarity.
\textbf{3) The importance of accounting for temporal information.} 
All time-sensitive measures outperform their non-temporal counterparts (e.g. t-LR vs. LR, or t-SED vs. SED). 
This result implies significant amounts of concept drifting in the dataset. 
We also note that t-ED and t-SED have similar performances, despite the performance difference between ED and SED.
This seems to suggest that the temporal information is more important for the successful classification than the word similarity. 
However, the best value for $k$ obtained on the validation set shows that t-SED can use a larger number of neighbours than t-ED, which implies the potential robustness of t-SED against t-ED by accounting wider context through the word similarity.
\textbf{4) Circumvention of trainable models.} 
Given the most accessible features, the trace of text, t-SED outperforms training based models (i.e. LR and t-LR). 
This implies that distance-based methods can outperform training models in the most restricted scenario where only the authored text is available (without user attributes). 
It is also important to note that the distance-based methods do not need to be retrained, whereas the training based models, such as the logistic regression, require periodical retraining to model temporal changes in a corpus.

Note that the two normalised metrics, SED/Max and SED/ED, do not help increase the performance. We conjecture that a normalisation is unnecessary due to the limited number of characters can be written at a time\footnote{Twitter has a 140-character limitation before Nov. 2017}.


\section{Results and findings: the strategy of trolls}
\label{sec:visualisation}

\begin{figure*}[t!]
    \centering
    \begin{subfigure}{0.48\linewidth}
    \includegraphics[width=\linewidth]{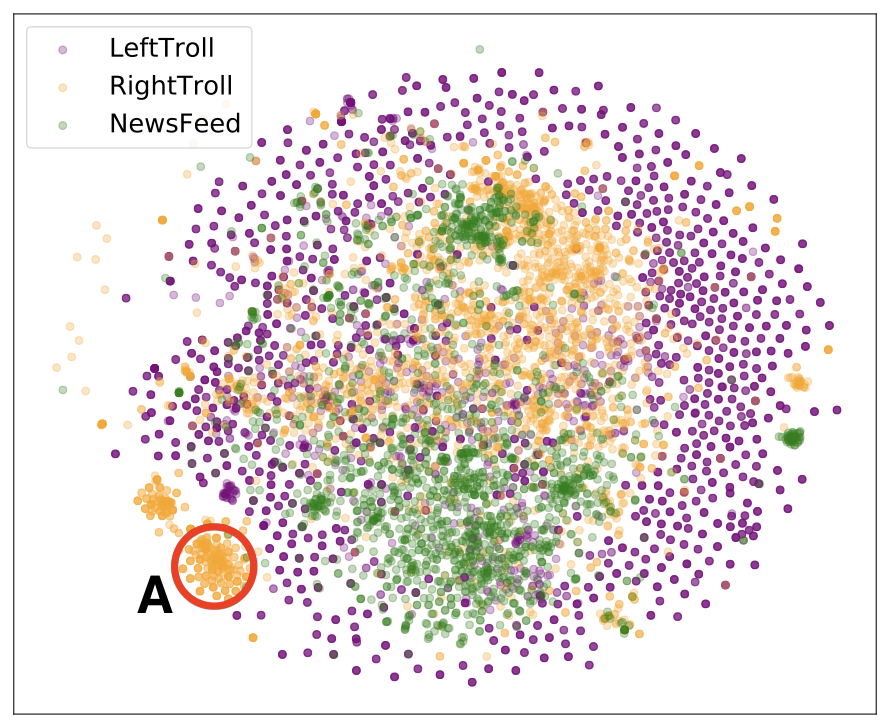}
    \caption{[SED] September, 2016}
    \label{fig:sed-sep2016}
    \end{subfigure}\hfill
    \begin{subfigure}{0.48\linewidth}
    \includegraphics[width=\linewidth]{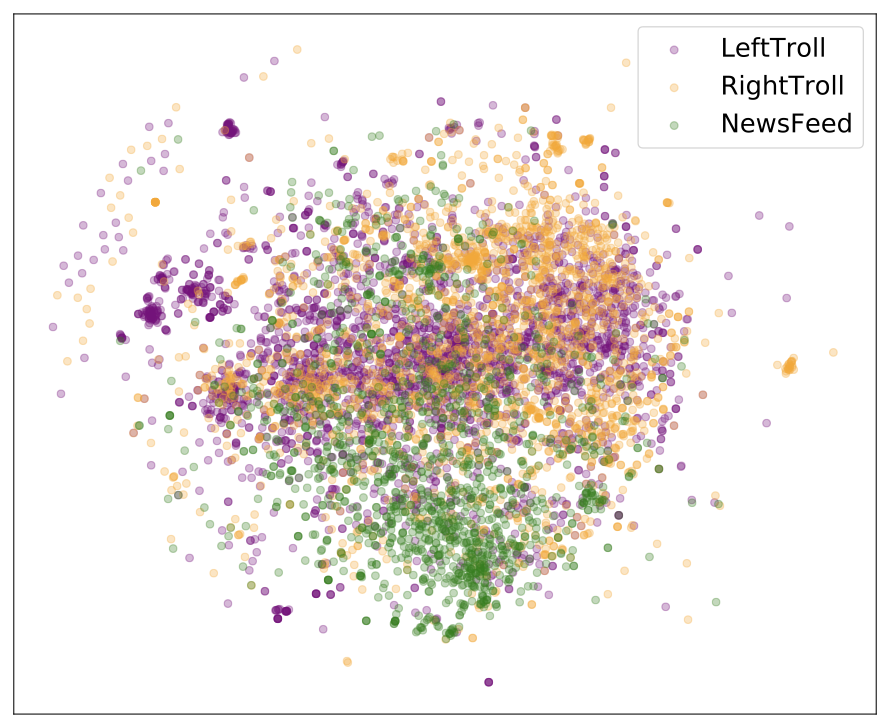}
    \caption{[SED] April, 2017}
    \label{fig:sed-apr2017}
    \end{subfigure}

    \begin{subfigure}{0.48\linewidth}
    \includegraphics[width=\linewidth]{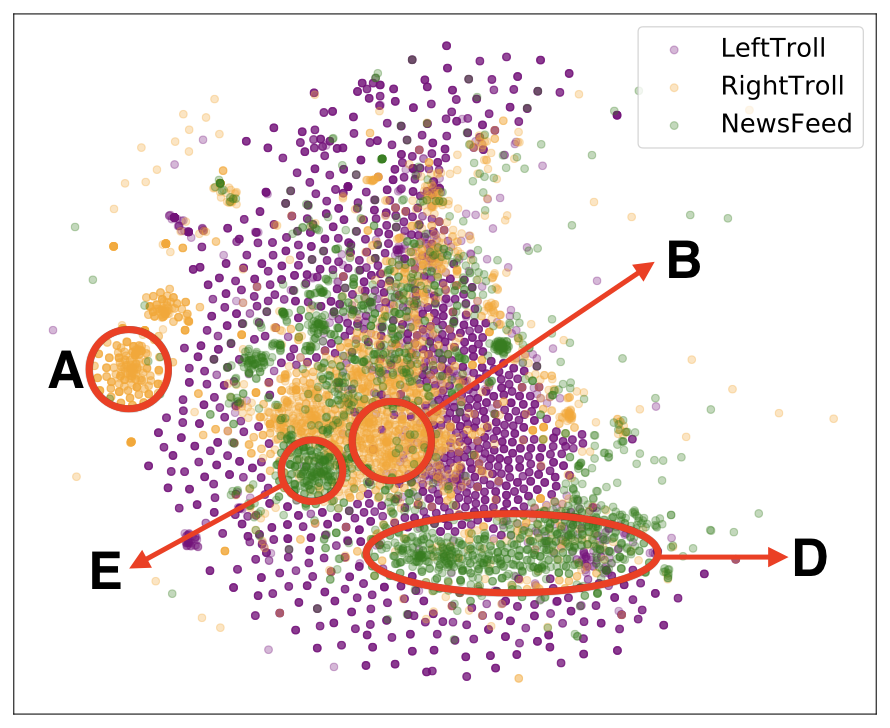}
    \caption{[t-SED] September, 2016}
    \label{fig:tsed-sep2016}
    \end{subfigure}\hfill
    \begin{subfigure}{0.48\linewidth}
    \includegraphics[width=\linewidth]{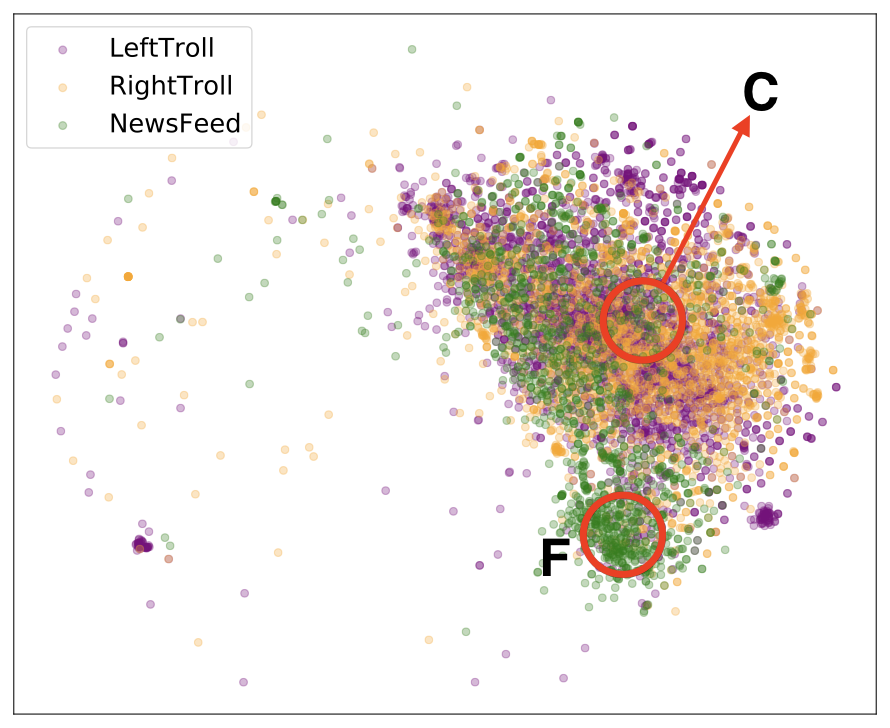}
    \caption{[t-SED] April, 2017}
    \label{fig:tsed-apr2017}
    \end{subfigure}
    \caption{
        Tweets from two different time ranges (left vs right column) are embedded into two-dimensional spaces, via t-SNE, with two variants of edit distances (SED for the top row, and t-SED for the bottom row). 
        The locations are computed using the distances between pairwise distances between all tweets from the two time ranges -- i.e. one space for SED and one for t-SED.
        We plot the tweets separately based on the period they were written. 
        For SED, there is a relatively clear distinction between the different categories before the election, but they are relatively indistinguishable after the election. 
        For t-SED, the gap between 2016 and 2017 is wider than for SED since the distance between tweets increases exponentially with their time difference.
    }
    \label{fig:embedding}
\end{figure*}

To understand the behaviour and the strategy of trolls over time, we visualise the tweets using t-SNE~\cite{maaten2008visualizing}, a technique originally designed to visualise high-dimensional data through their pairwise distances. 
\autoref{fig:embedding} shows the visualisation of tweets from two different time ranges using SED and t-SED.
Each dot in the figure represents a single tweet embedded into a two dimensional space constructed by t-SNE.
To emphasise different behaviour of trolls over time, we plot tweets from two different time ranges: one before the presidential election (September 2016) in \autoref{fig:sed-sep2016} and \autoref{fig:tsed-sep2016}, 
and another after the election (April 2017) in \autoref{fig:sed-apr2017} and \autoref{fig:tsed-apr2017}.

\subsection{Re-examination of Russian trolls}

\textbf{Together they troll.}
The Clemson researchers have argued that the troll handles are consistent in their roles (i.e. they don't switch between categories), and that they often talk about the same topic, but they do so in a different way and to different audiences. 
Our own observations lend further support to this finding -- a cursory examination of the visualisation shows several instances where the positioning of tweets from both right trolls and left trolls overlap, suggesting that they tweet about similar or related topics.
Similarly, \autoref{fig:sed-sep2016} and \autoref{fig:tsed-sep2016} reveal sub-clusters of news feed trolls that are positioned within the predominantly right troll cluster. 
Therefore, we cannot observe a clear separation in authored tweets based on their categories through the distance metrics.
Nonetheless, many tweets are locally clustered in line with their categories, which ultimately helps us to correctly classify their type, as shown in Section \ref{sec:result-section}.

\textbf{Right vs. left strategy.}
A notable pattern from \autoref{fig:embedding} is that the tweets authored by the left trolls are spread across all regions whereas those of the right trolls are more focused and relatively well clustered. 
As discussed previously, we know that generally the right trolls were focused on supporting Trump and criticising mainstream Republicanism and moderate Republican politicians. 
Compared to left trolls, right trolls have a more singular or homogeneous identity, and employ common hashtags used by similar real Twitter users, including \#tcot, \#ccot, and \#RedNationRising \citep[p. 7]{boatwrighttroll}. 
On the other hand, left trolls have a more complex discursive strategy. 
As \citep{boatwrighttroll} argue, these trolls send socially liberal messages, with an overwhelming focus on cultural identity.
Accordingly, the position of the left trolls on the visualisation provides a stark picture of their complementary and supporting role in driving the IRA's agenda building campaign on Twitter. 
Left trolls are literally surrounding the conversation on all sides. 
In some areas they are attacking Hillary Clinton and mainstream Democratic politicians, in others they are mimicking activists from the Black Lives Matter movement, and in others discussing religious identity and Christian moralism. 
Left troll tweets are certainly distinguishable on the visualisation in terms of their position, but we can see how they play into, and strategically function alongside, the news feed and right trolls. 
To examine these observations in more detail we can zoom in to specific regions of interest within the visualisation to analyse the tweet content.

\subsection{Left and right trolls worked together}

\textbf{Leveraging racial politics.}
Most of the tweets from the notable right troll cluster in \autoref{fig:sed-sep2016} and \autoref{fig:tsed-sep2016} (\textbf{tag A} in the figures) contain the hashtag \#ThingsMoreTrustedThanHillary, which shows the strategic behaviour of certain right trolls to make the Democratic candidate distrustful. 
This strategy is also part of the over-arching agenda of the left trolls, who not only undermined the trust in Hillary Clinton, but co-opted the Black Lives Matter movement and related topics to negatively impact her campaign. 
As \citep[p. 8]{boatwrighttroll} show, left trolls authored tweets such as ``NO LIVES MATTER TO HILLARY CLINTON. ONLY VOTES MATTER TO HILLARY CLINTON'' (@Blacktivists, October 31, 2016). 
Furthermore, the central cluster of right trolls (\textbf{tag B}) in \autoref{fig:tsed-sep2016} contains tweets that show the support of Trump from black people, in addition to tweets from typical Trump supporters. 
The following are example tweets from this cluster of right trolls:
\begin{itemize}
    \item { ``Join Black Americans For Trump, ``Trump is the best choice for ALL Americans!'' Join Today at https://t.co/NJBoTamxDi  \#Blacks4Trump'' (@marissaimstrong)};
    \item { ``Why I Support Donald Trump https://t.co/U0oT8odMOB \\ \#BlacksForTrump \#Blacks4Trump \#BlackLivesMatter \#ImWithHer \#DemExit \#MAGA'' (@hyddrox)}.
\end{itemize}
We therefore observe a complex interplay between left and right trolls, whereby both attempt to co-opt and strategically utilise racial politics and ethnic identity \citep{phinney2000ethnic},
even though they use different approaches. 
This resonates with and provides new insights on recent analysis of Russian troll communication on Twitter, where trolls were found to make a ``calculated entry into domestic issues with the intent to polarise and destabilize'' \citep[p. 4]{stewart2018examining}.

\textbf{Utilising religious beliefs.}
From the central part of \autoref{fig:tsed-apr2017}, we observe tweets from both ideologies and we find a cluster of conversations (\textbf{tag C}) related to personal religious beliefs, such as:
\begin{itemize}
    \item { ``Just wait \& \#God will make things great'' (@acejinev)};
    \item { ``Each of us is a Masterpiece of God’s Creation. Respect Life! \#Prolife'' (@j0hnlarsen)}.
\end{itemize}
Although the hashtags used in these religious tweets are often different for the left and right trolls, their similarity is captured by the metric.
This reveals an interesting strategy whereby trolls from both left and right pretend to be ordinary American citizens who, although ideologically different, are united by shared religious beliefs. 
This indicates that not all trolls in a category acted unitarily, and the tweets they emitted cluster into groups corresponding to their different sub-roles and strategies.
Using the proposed t-SED measure and visualisation, one can zoom in and gain richer insights into the strategies and identities of these user accounts (down to individual actions).

\subsection{The multiple agendas of news feed trolls}
When studying news feed trolls, we observe how different clusters promote specific themes of news articles. 
The slender cluster of news feed trolls (\textbf{tag D}) in \autoref{fig:tsed-sep2016} often contain the hashtag \#news, and report incidences of violence and civil unrest. For example, tweets from this cluster include:
\begin{itemize}
    \item { ``Warplanes hit Aleppo in heaviest attack in months, defy U.S.  \#news'' (@specialaffair)};
    \item { ``Pedestrian hit, killed by train in Mansfield https://t.co/kvmFEgf8Ps \\ \#news https://t.co/TXsol3YjgA'' (@todaybostonma)}; 
    \item { ``One person killed during violent Charlotte protest; officer hurt \\ https://t.co/IYyg0xmf0L https://t.co/UbzzAeW3zR'' (@baltimore0nline)}.
\end{itemize}
On the other hand, the small left-most cluster of news feed trolls (\textbf{tag E}) in \autoref{fig:tsed-sep2016} focus on the hashtag \#politics, and have a focus on federal political issues and politicians as well as policy and regulation. Tweets from this cluster include, for example:
\begin{itemize}
    \item { ``Obama Trade Setbacks Undercut Progress in Southeast Asian Ties \\  \#politics'' (@newspeakdaily)};
    \item { ``Is federal government trying to take down for-profit colleges?  \#politics'' (@batonrougevoice)};
    \item { ``Clinton takes aim at Trump supporters https://t.co/fxEox7N74Z \#politics'' (@kansasdailynews)}.
\end{itemize}
We observe that the clusters help to illuminate the within-group variation for this troll category, and we might speculate that the clusters correspond to, or at least highlight, the agenda-setting strategies that news feed trolls carried out, as well as their relationship to other types of trolls (i.e., by analysing their proximities in t-SNE space).

\subsection{The shifting sands of troll identity over time}
By analysing two different time ranges, we can notice that there is less separation between the tweets belonging to the different roles
(see \autoref{fig:sed-apr2017} and \autoref{fig:tsed-apr2017}). 
The clustering structure around September 2016 is comparatively clearer than the structure around April 2017. 
The difference suggests that, for some reason, the strategic behaviour of trolls changed markedly before and after the election. 
We are not aware of any previous research that has identified this, let alone that offers an explanation why.
Interestingly, one strategy that appears to continue after the elections is seeding fear: the t-SED visualisation in \autoref{fig:tsed-apr2017} reveals a cluster of news feed trolls (\textbf{tag F}) with a particular focus on reporting negative news about shootings and gun violence, crime and fatalities. 
Example tweets from this cluster include: 
\begin{itemize}
    \item { ``Gunman shoots woman at Topeka Dollar General \\ https://t.co/gQgQy8B0Hh'' (@kansasdailynews)};
    \item { ``Police: Father accidentally shoots son while fighting off intruder \\ https://t.co/kAD9shfp7t'' (@kansasdailynews)}; 
    \item { ``Police: Suspect dead after woman, child abducted in Homewood \\ https://t.co/TcTNmc5oFu'' (@todaypittsburgh)}.
\end{itemize}
Although we do not have scope in this paper to undertake further analysis of the clustering after the election, it is obvious that t-SED (\autoref{fig:tsed-apr2017}) offers a different view of Russian troll strategies as compared to SED (\autoref{fig:sed-apr2017}).
Zooming out to the largest scale, we see generally that taking into account temporal information is important because it outlines the drift in topics of discussion.

\section{Discussion}
\label{sec:discussion}

As we have shown, the visualisation and analysis presented in Section~\ref{sec:visualisation} affords a nuanced analysis of the \textit{overlap} and heterogeneity of Russian troll identities and strategies, which are not as disjoint or homogeneous as previous work suggested. This is not to say that prevailing analytic categories of Russian trolls are insufficient or invalid -- on the contrary, what we offer here builds upon and extends existing scholarship. Applying a sequence analysis approach to this problem not only coherently recovers the identity labels from previous work (left, right and news feed trolls), it also discovers new aspects of the troll identities and strategies, which in turn complements and enhances our understanding of them. Fundamentally, however, we believe that the framework we have developed and evaluated in this paper has significant relevance for new frontiers in computational social science, based on developments in the field of Actor-Network Theory (ANT). 

\textbf{Revisiting how social order is generated}. 
Gabriel Tarde's ancient theory of monadology~\citep{tarde2011monadology} has recently been adapted into ANT.
It promises a powerful framework for the study of identity and social change in heterogeneous networks \citep{latour2012whole}. 
In the 19$^{th}$ century, Tarde's ideas proved not only difficult to conceptualise but even more difficult to operationalise due to a lack of data. It is perhaps for this reason that his alternative approach to describing social processes was not empirically testable and subsequently relegated to a footnote in history.

However, \citet{latour2012whole} argue that the onset of the information age and the availability of digital data sets make it possible to revisit Tarde’s ideas and render them operational. 
By examining the digital traces left behind by actors in a network (human and non-human), \citet[p.~598]{latour2012whole} argue that we can `slowly learn about what an entity ``is'' by adding more and more items to its profile'. 
The radical conclusion is that datasets `allow entities to be individualised by the never-ending list of particulars that make them up' \citep[p.~600]{latour2012whole}. There is only one level of reality to deal with: `society' and the actors that constitute it all interact and exist at the micro level. This idea goes against conventional thinking in social science, where a distinction is often made (explicitly or implicitly) between two levels, micro and macro, with a third `meso' level sometimes added in between. Not surprisingly, the nascent field of computational social science also inherits this assumption about society and its generative processes. In the traditional sense, actions begin at the micro level and somehow filter up to the macro level where they form `structures', and in some cases filter back down again in a feedback loop. Yet for \citet{latour2012whole} and ANT, there is no macro level of society. An entity can be fully understood by tracing its activities through data and comparing the similarities and differences with the traces of other entities \textit{on the same level}. In other words, we can define an actor through the network of other entities it is attached to, hence its `actor-network'.

\textbf{Rethinking the `social' in computational social science}. 
Based on the framework and findings of this paper, we suggest that this idea could be usefully envisioned as a sequence analysis problem. In this way, a simple interpretation of an actor-network is an ordered sequence of events relating to, or initiated by, a particular actor. 
For Twitter, these events are tweets defined by their timestamps, words and hashtags. 
In this paper we have shown how words and hashtags are unique symbols that can be represented in a trace sequence: as we traverse linearly along a particular users' trace sequence of tweets, we start to `zero in' on their specific identity as expressed through what they write about. 
Rather than pre-constructing macro-level or structural attributes for actors, such as `ideology' or `gender', we can coherently derive such characteristics by analysing and comparing similarities of the sequences of traces they leave behind (see Section \ref{sec:sed}).

From a sequence analysis point of view, each time that we wish to pinpoint the identity or social role of a Russian troll, we can look to the elements of its trace sequence (in this case tweets, but also potentially location, meta-data, etc). 
What matters are the similarities in sequences that pass between actors
from one time point to another \textit{on the same level}. This affords an analysis of the `partial wholes' of social action and identity, rather than fully-formed wholes that exist on different levels of reality (e.g., individual node versus aggregate structure) and somehow must interact with one another whilst also being ontologically distinct. There are not multiple `levels' of reality (e.g. micro and macro) -- instead, we have a one-level standpoint whereby actors (e.g. Russian trolls) are defined by their network of traces, which have differing degrees of similarity with other traces that we can quantify via edit distance. As \citep[p. 593]{latour2012whole} write: ``This network is not a second level added to that of the individual, but exactly the same level differently deployed''.

The work we have presented in this paper suggests that it is possible to formulate an understanding of social identity and social roles through visualising and analysing the 2D plane in which the similarity between trace sequences are arranged. From a geometric point of view, this is perhaps somewhat literally the `one-level standpoint' that \citep{latour2012whole} have argued for. By incorporating a semantic cost and temporal cost into edit distance (via t-SED), we see that the closer two entities are together in the 2D visualisation (e.g., two tweets), the more their constituent elements repeat with similar variation over time (e.g., words and hashtags that appear at similar time periods).


To be sure, we do not wish to overstate the usefulness of our methods and findings in this paper. Despite the apparent relevance to a longstanding problem in social theory, a glaring criticism is that the results of our approach simply served to rediscover, or rather predict, the Russian troll labels (left, right, news feed), which we could regard as the very sort of a priori structures that ANT rejects. In a sense, we end up getting right back to structure. However, more than simply recovering the a priori labels, we `rediscover' them. A sequence analysis approach helps us to discover, a posteriori, interesting variations and nuances that constitute the over-arching labels. It helps us to quantify and decompose the types of Russian trolls according to the constituent elements of their traces. Following this, we can compare and contrast the similarities and differences of these traces, and in doing so retrieve valid and sensical results (that should accord with the known labels) as well as gaining new insights and knowledge about the social phenomenon under examination, in this case Russian trolls on Twitter.

\section{Conclusion}

In this study, we developed a novel framework to map and analyse user identity and social roles in social media data. 
We formalised this idea computationally using social sequence analysis \citep{abbott2000sequence} and as a data clustering problem. 

To develop and evaluate this method, we addressed a new challenging case study: how to characterise online trolls and understand their tactics based on their social roles and strategies. We focused on the different types of trolls identified in recent studies to picture a more detailed - and theoretically nuanced - landscape of how Russian trolls attempted to manipulate public opinion and set the agenda during the 2016 U.S. Presidential Election. 

We define a novel text distance metric, called \textit{time-sensitive semantic edit distance} (t-SED), and we show the effectiveness of the new metric through the classification of Russian trolls against ground-truth labels (left-leaning, right-leaning, and news feed trolls).
The metric is then used to construct a novel visualisation for qualitative analysis of troll identities, social roles and strategies. Through the application of t-SED and our framework for analysing identity and social action, we discover intriguing patterns in the similarities of traces that Russian trolls left behind via their tweets, providing unprecedented insights into Russian troll activity during and after the election. 

We believe the results of this paper constitute a promising contribution to developing social theory within computational social science. As Cornwell writes, `the social sciences are full of well-theorized but seldom-tested ideas about the structural causes and consequences of the ordering of social events' \citep[p.~xvii, preface]{cornwell2015social}. Hence this paper identifies a point of connection between Actor-Network Theory and social sequence analysis, which we hope will advance both fields and open up new avenues of research in computational social science.

\textbf{Assumptions, limitations and future work.}
This work makes a number of simplifying assumptions, some of which can be addressed in future work. 
First, for the empirical analysis we assume that each tweet is assigned exactly one label, selected from the same set as the user labels. 
Future work will relax this assumption, allowing tweets to have multiple labels, possibly from a set disjoint from the user labels.
Second, we measure the similarity between the traces of two users by measuring the similarity between tweets and performing a majority vote. This could be extended by introducing similarity metrics directly working on a trace level instead of using an aggregated approach.
Third, we have identified a promising avenue for combining social sequence analysis and the field of Actor-Network Theory. Future work may develop further links between social theory and the computational methods we set out and evaluated in this paper. In doing so, this will contribute to developing empirically useful computational tools that also incorporate and exploit the explanatory power of social theory.
Fourth, future studies may apply this method to other kinds of digital trace data, as it is unclear from the case study in this paper whether, and to what extent, the method will elicit novel insights into other social phenomena. For example, we envisage that the methods could be particularly useful for tracking the propagation and evolution of memes in social media, the unfolding of citation networks and academic identities, and of course further understanding misinformation campaigns involving bots and trolls.
Finally, we aim to construct and publish an interactive version of the visualisation in \autoref{fig:embedding}.



\section{Conflict of Interest}
On behalf of all authors, the corresponding author states that there is no conflict of interest.

\bibliographystyle{spbasic}      
\bibliography{paper.bib}


%
%

\end{document}